\documentclass[showpacs,prl,reprint,amsmath,amssymb,superscriptaddress,nobibnotes,twocolumn, preprintnumbers, bibnotes]{revtex4-1}
\usepackage{graphicx}
\usepackage{endnotes}
\usepackage{bm}
\usepackage{epsfig}
\usepackage{amsmath}
\usepackage{color}
\usepackage{xcolor}
\usepackage{array}
\usepackage[bottom,flushmargin,hang,multiple]{footmisc}

\newcommand{\parallelsum}{\mathbin{\!/\mkern-5mu/\!}}

\begin{document}


\title{Extremely large magnetoresistance and the complete determination of the Fermi surface topology in the semimetal ScSb}

\author{Y.~J.~Hu$^\ddagger$}
\author{E.~I.~Paredes~Aulestia$^\ddagger$}
\author{K.~F.~Tse}
\affiliation{Department of Physics, The Chinese University of Hong Kong, Shatin, New Territories, Hong Kong, China}

\author{C.~N.~Kuo}
\affiliation{Department of Physics, National Cheng Kung University, Tainan 70101, Taiwan}

\author{J.~Y.~Zhu}
\affiliation{Department of Physics, The Chinese University of Hong Kong, Shatin, New Territories, Hong Kong, China}

\author{C.~S.~Lue}
\affiliation{Department of Physics, National Cheng Kung University, Tainan 70101, Taiwan}

\author{K.~T.~Lai} 
\affiliation{Department of Physics, The Chinese University of Hong Kong, Shatin, New Territories, Hong Kong, China}

\author{Swee~K.~Goh}
\email{skgoh@phy.cuhk.edu.hk}
\affiliation{Department of Physics, The Chinese University of Hong Kong, Shatin, New Territories, Hong Kong, China}
\affiliation{Shenzhen Research Institute, The Chinese University of Hong Kong, Shatin, New Territories, Hong Kong, China}

\date{\today}


\begin{abstract}
We report the magnetoresistance of ScSb, which is a semimetal with a simple rocksalt-type structure. We found that the magnetoresistance reaches $\sim$28000\% at 2~K and 14~T in our best sample, and it exhibits a resistivity plateau at low temperatures. The Shubnikov-de Haas oscillations extracted from the magnetoresistance data allow the full construction of the Fermi surface, including the so-called $\alpha_3$ pocket which has been missing in other closely related monoantimonides, and an additional hole pocket centered at $\Gamma$. The electron concentration ($n$) and the hole concentration ($p$) are extracted from our analysis, which indicate that ScSb is a nearly compensated semimetal with $n/p\approx0.93$. The calculated band structure indicates the absence of a band inversion, and the large magnetoresistance in ScSb can be attributed to the nearly perfect compensation of electrons and holes, despite the existence of the additional hole pocket.

\end{abstract}


\maketitle

\section{I. Introduction}

Magnetoresistance (MR), the change of the electrical resistance when a magnetic field is applied, is usually a weak effect in conventional nonmagnetic metals \cite{Pippard}. Hence, materials with a large MR naturally attract intense attention concerning the underlying fundamental physics as well as their potential technological impact. The importance of the MR magnitude is fully reflected in the coining of special terms such as {\it giant} MR (GMR) \cite{Baibich1988} and {\it colossal} MR (CMR) \cite{Shapira1973,Jin1994}. Recently, very large MR has been reported in several topological semimetals such as WTe$_2$ \cite{Ali2014,Wang2015,Jiang2015, Luo2015} and NbAs$_2$ \cite{Wang2016A,Yuan2016,Wu2016B,Luo2016,Shen2016} and PtSn$_4$ \cite{Mun2012}, thereby expanding the material base for the {\it extremely large} MR (XMR) family. In addition to the magnitude of the MR, the way the MR changes in the magnetic field has also attracted significant research efforts: in some cases the MR varies linearly in the magnetic field (e.g. Refs.~\cite{Narayanan2015, Niu2017}), while in others it is a quadratic function of the magnetic field (e.g. Refs.~\cite{Ali2014,Wang2015,Jiang2015, Wang2016A, Mun2012}). Thus, the measurement of materials showing an interesting MR behaviour supported by a thorough electronic structure investigation is a promising route to enrich the understanding of these materials.

Recently, rare-earth monopnictides $RX$ ($R$=rare earth, $X$=Sb, Bi) with the rocksalt-type structure have drawn attention because they also exhibit the XMR behaviour \cite{Tafti2016B,Sun2016,Kumar2016,Singha2017,Niu2016,Ghimire2016,Yu2017,Pavlosiuk2016,He2016,Xu2017,Tafti2016A,Zeng2016,Ban2017,Han2017,Wakeham2016,Pavlosiuk2017A,Dey2018,Wu2017C,Wang2018,Pavlosiuk2017B,Song2018,Ye2018}. An important member of the family is LaBi \cite{Tafti2016B,Sun2016,Kumar2016,Singha2017,Niu2016}, which has been proposed to be a topologically nontrivial semimetal \cite{Zeng2015}. In LaBi, several intriguing magnetotransport properties have been reported: (i) a large, nonsaturating magnetoresistance reaching 1.4 $\times 10^{5}$
 \% at 9~T and 2~K, (ii) a field-induced upturn in the resistivity, and (iii) a resistivity plateau at high field and low temperature, analogous to the observation in SmB$_6$ at zero field \cite{Kim2013,Kim2014}. These unusual transport properties in LaBi have been attributed to the surface states and the band inversion, or orbital texturing, near the X point of the Brillouin zone \cite{Tafti2016B}. Similar transport properties have also been observed in YSb \cite{Ghimire2016,Yu2017,Pavlosiuk2016,He2016,Xu2017} and LaSb \cite{Tafti2016A,Zeng2016,Ban2017,Han2017,Niu2016}, which are both isostructural to LaBi. However, angle-resolved photoemission spectroscopy (ARPES) did not detect any nontrivial topological state in both YSb \cite{He2016} and LaSb \cite{Zeng2016}. Furthermore, Kohler scaling was performed to elucidate the origin of the resistivity plateau and the field-induced upturn in the resistivity, ruling out the role of the surface states on the magnetotransport properties \cite{Han2017,Pavlosiuk2016}.

ScSb, isostructural to LaBi and YSb, is the first member of the rare-earth monoantimonide ($R$Sb) series. Although the Fermi surface topology is qualitatively similar to that of LaBi and YSb, two important differences arise in ScSb upon inspecting the calculated bandstructures in literature \cite{Maachou2007, Xue2009}. First, there seems to be an additional hole pocket centered at the $\Gamma$ point of the Brillouin zone. Second, there is a large energy gap between the band formed by Sc $d$ states and Sb $p$ states near the X point of the Brillouin zone. Thus, band inversion is absent. In this manuscript, we present the MR and the Shubnikov-de Haas (SdH) oscillations of single-crystalline ScSb, supported by density functional theory (DFT) calculations. Despite the key differences in the band structure, which are confirmed by our own calculations, our MR curves are remarkably similar to the cases where nontrivial topological states have been asserted. Furthermore, our SdH oscillations detect all Fermi surface pockets predicted by calculations. Therefore, ScSb serves as a useful reference compound in which the bulk Fermi surface topology is fully determined to distill the essential ingredients responsible for the interesting magnetotransport properties in rare-earth monoantimonides.

\section{II. Experimental}

\begin{figure}[!t]\centering
      \resizebox{8.5cm}{!}{
              \includegraphics{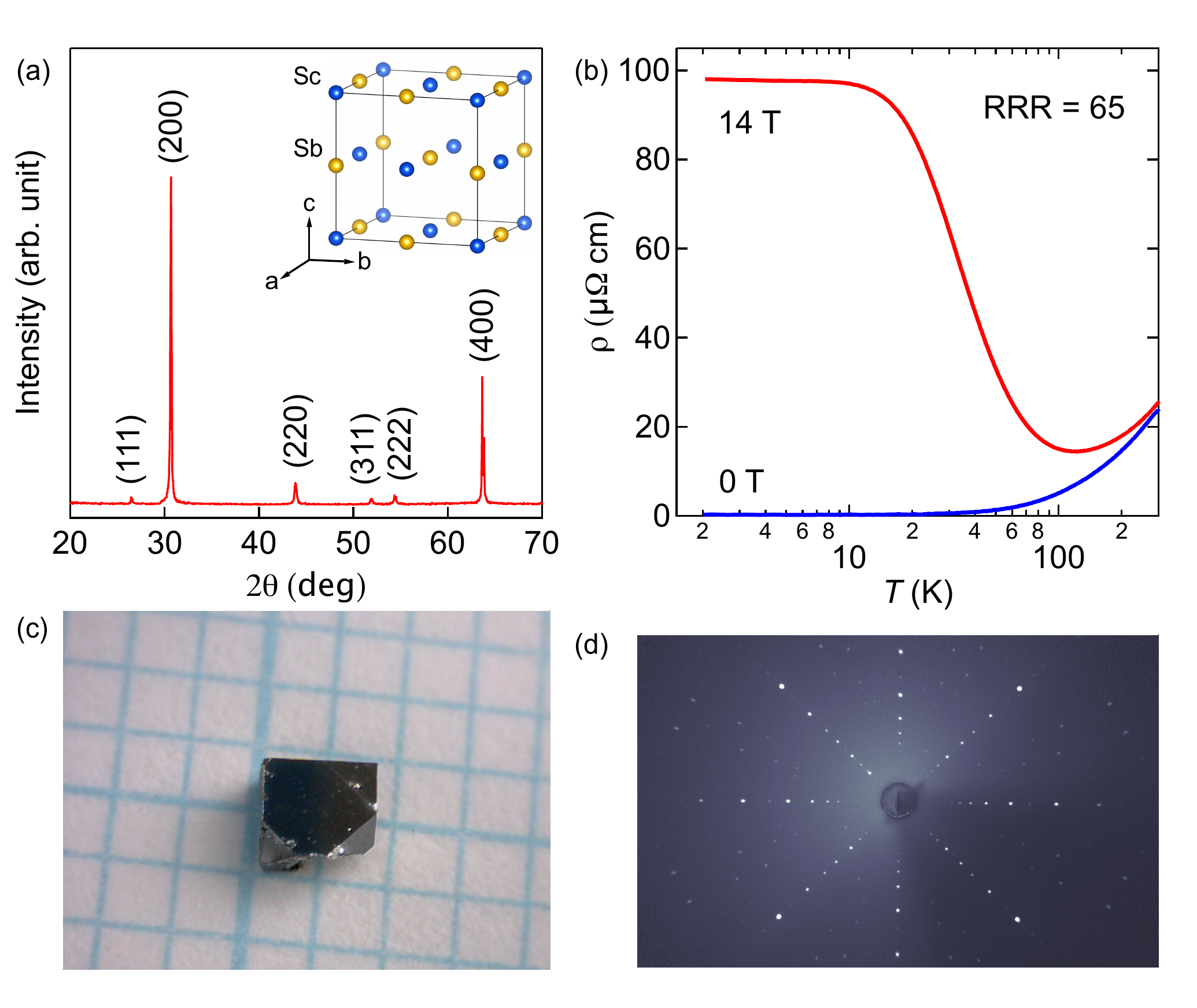}}                				
            \caption{\label{fig1}(a) Powder X-ray diffraction pattern of ScSb. The crystal structure of ScSb is shown in the inset. (b) Temperature dependence of the electrical resistivity for the ScSb crystal with RRR of 65 (SS21) at 0 T and 14 T. (c) Photograph of a typical ScSb single crystal (each grid stands for $1\times 1$ mm$^2$). (d) Laue diffraction image taken along the [100] direction.}
\end{figure}

Single crystals of ScSb were grown in Sb self-flux. High-purity Sc (99.99\%) and
Sb (99.999\%) were mixed with a molar ratio between 1:5 and 1:15. The mixture was then loaded into alumina crucibles
and sealed under vacuum in a quartz tube. The sealed quartz tube was heated to
1050 $^\circ$C, and maintained at 1050 $^\circ$C for 8 hours before cooling to 750
$^\circ$C at a rate of 3--5$^\circ$C/h. The excess Sb was removed using a centrifuge. 
The crystal structure and phase purity of the as-grown crystals were
examined by X-ray diffraction (Bruker D2 PHASER) using Cu K$\alpha$ radiation and
Laue diffraction (Photonic Science) at room temperature. The electrical resistivity was measured using a four-terminal configuration. Low-temperature (down to 2~K) and high magnetic field (up to 14~T) environments were provided by a Physical Property Measurement System (PPMS) made by Quantum Design. A motorized rotator was employed to vary the orientation of the crystal with respect to the magnetic field direction. Throughout the entire measurement, the current direction was perpendicular to the magnetic field direction. 
Density functional theory (DFT) calculations were performed using the projector augmented wave (PAW) method within the meta-generalized gradient approximation (mGGA) including spin-orbit coupling (SOC) as implemented in Vienna Ab initio Simulation Package (VASP) \cite{Kresse1996A}. The Strongly Constrained Appropriately Normed (SCAN) functional \cite{Sun2015} was used as the approximation to exchange and correlation energy. Further computational details, such as the convergence criteria, as well as comparisons with HSE06 functional \cite{Heyd2003} and GGA functional \cite{Perdew1996} are presented in Supplemental Material \cite{SUPP}. 
 
\section{III. Results and Discussion}
\begin{figure}[!t]\centering
       \resizebox{8.5cm}{!}{
\includegraphics{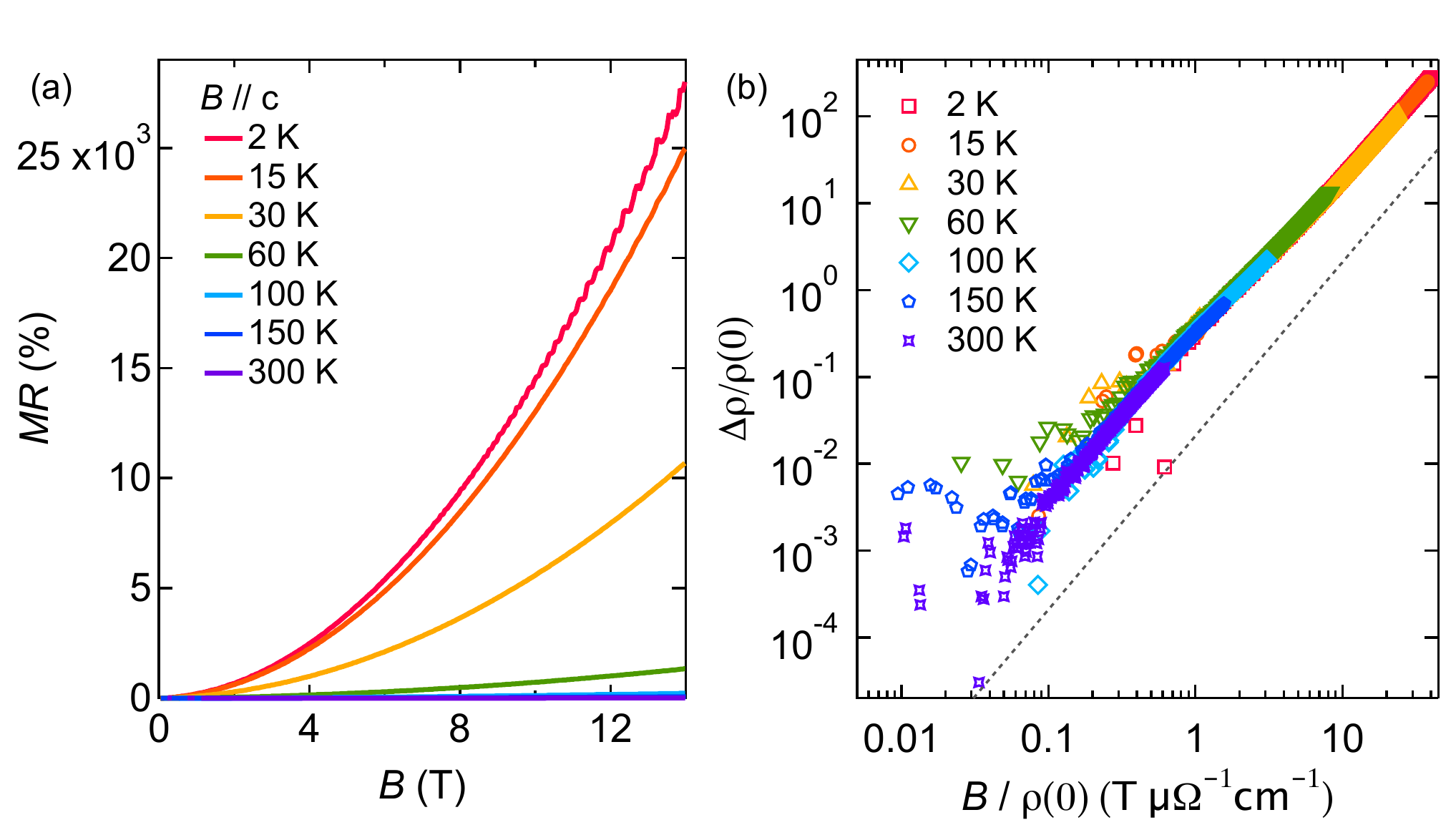}}                				
\caption{\label{fig2} (a) Magnetoresistance with the field applied along the $c$-axis at different temperatures. (b) Magnetoresistance in (a) plotted against $B/\rho(0)$, where $\rho(0)$ is the resistivity at zero field. Note that logarithmic scales are used. The dashed line indicates a slope of 2 on the log-log plot.}
\end{figure}

Figure~\ref{fig1}(a) shows a typical X-ray diffraction pattern of powdered ScSb. The refinement of the powder pattern confirms the rocksalt structure of ScSb (space group: $Fm\bar{3}m$) as displayed in the inset of Fig.~\ref{fig1}(a), and gives the lattice constant $a=$~5.8336~\AA. The lattice constant is the smallest among the $R$Sb series, which is not surprising because Sc is the first and the smallest element of the series. The sharp lines in the diffraction pattern indicate the high crystal quality of our sample. Figure~\ref{fig1}(c) shows the photograph of a representative as-grown ScSb single crystal. The Laue diffraction was performed on the single crystal, and the diffraction pattern along the [100] direction is displayed in Fig.~\ref{fig1}(d). These well-defined spots shown on the pattern prove that our ScSb crystals are highly single-crystalline.
The temperature dependence of the electrical resistivity ($\rho(T)$) was measured for 10 crystals from several batches at 0~T, from which the residual resistivity ratio (RRR$=\rho(300~{\rm K})/\rho(2~{\rm K})$) is calculated and showed in Section~S1 of the Supplemental Material \cite{SUPP}. Figure~\ref{fig1}(b) displays $\rho(T)$ for ScSb (sample number SS21) with RRR=65, the largest among the 10 samples measured. The zero field resistivity data exhibit a clear metallic behaviour with a very low residual resistivity of 0.35~$\mu\Omega$cm. With the application of $B=14$~T along the $c$-axis, $\rho(T)$ experiences a large enhancement below 120~K, before reaching a resistivity plateau below $\sim$~10~K. The magnetoresistance (MR), defined as $[\rho(B)-\rho(0)]/\rho(0)\times 100\%$, is 28000\% at 14~T and 2~K. A qualitatively similar $\rho(T, B)$ is observed in a medium quality ScSb sample (SS02, RRR=55), as discussed in the Supplemental Material \cite{SUPP}. Thus, $\rho(T, B)$ in ScSb exhibits a very similar behaviour as other monopnictides with the same structure.

Figure~\ref{fig2}(a) shows the field dependence of the MR at different temperatures. The MR is nonsaturating and increases as $B^n$ with $n=1.94$. Shubnikov-de Haas (SdH) oscillations are visible in the MR at 2~K and high field even without subtracting the background, and detailed discussions will be provided below. The MR collected over a wide temperature range can be collapsed onto a universal curve when it is replotted against $B/\rho(B=0)$ (Figure~\ref{fig2}(b)), satisfying the Kohler's rule and signifying a universal scattering mechanism from 2~K to 300~K. Following the Kohler analysis and the related argument presented for WTe$_2$ and LaSb \cite{Wang2015, Han2017}, we conclude that both the field-induced resistivity upturn and plateau come from the bulk states.
\begin{figure}[!t]\centering
       \resizebox{8.0cm}{!}{
              \includegraphics{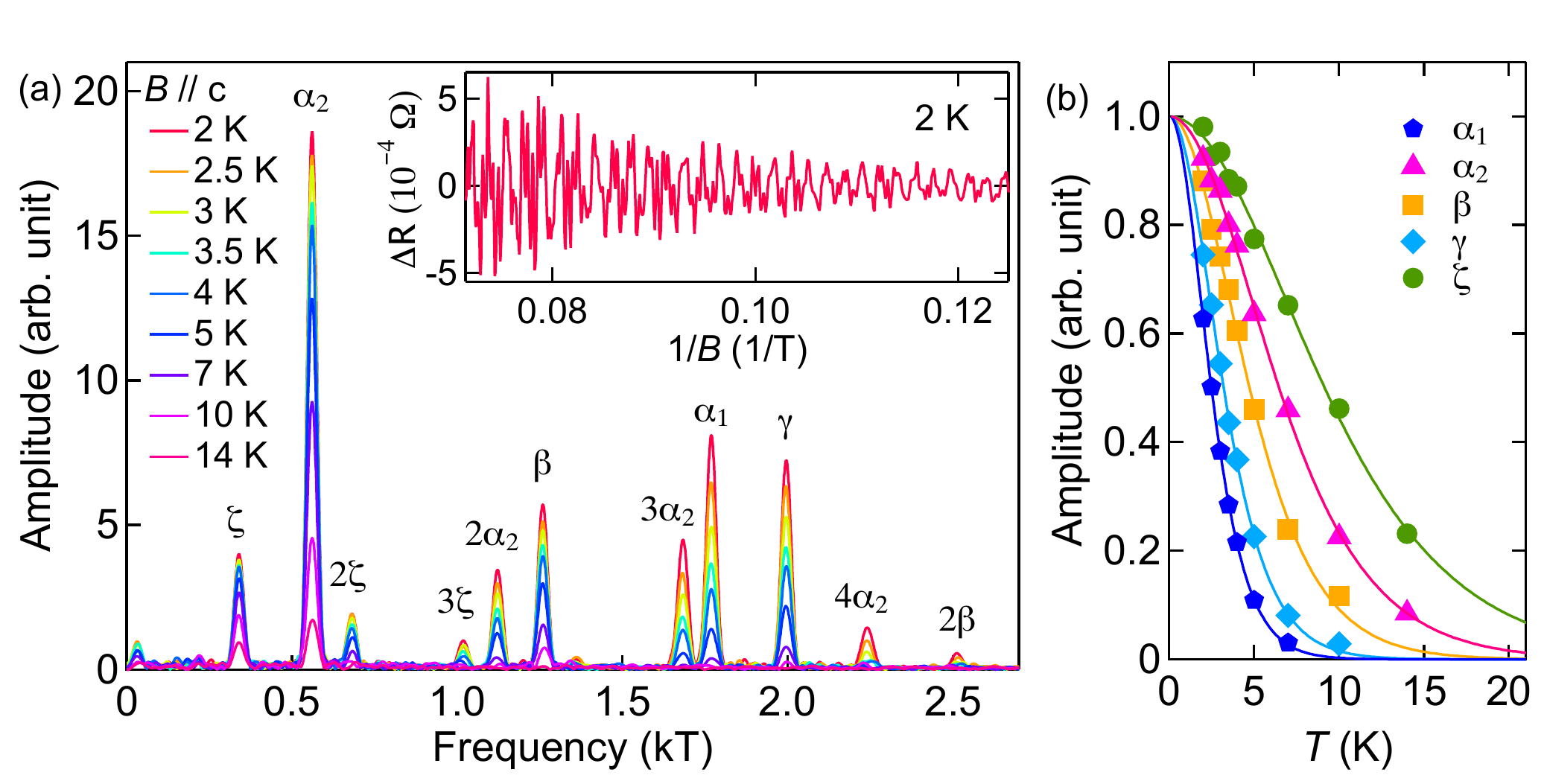}}
              \caption{\label{fig3} (a) FFT spectra of SdH quantum oscillations at different temperatures with $B\parallelsum c$. A representative dataset, with the steep MR background removed, is shown in the inset. (b) Temperature dependence of the oscillation amplitude for the fundamental frequencies, analyzed using the thermal damping factor (Equation~\ref{LKeq}) for the extraction of the effective masses.}
\end{figure}

\begin{table}[b]
\centering
\caption{Sheet-resolved effective masses with $B\parallelsum c$}
\label{my-label}
\begin{tabular}{clclclclclclc}
\hline
\hline
  && $\alpha_1$ && $\alpha_2$ && $\beta$ && $\gamma$ && $\zeta$ \\
 \hline
\textit{F} (T) && 1769  && 560  && 1258 && 1995 && 340 \\
$m^*/m_0$ && 0.61&& 0.23  && 0.32 && 0.50 && 0.16 \\
\hline
\end{tabular}
\end{table}

We now present the determination of the bulk Fermi surface via the SdH effect. The inset of Fig.~\ref{fig3}(a) shows the oscillatory resistivity at 2~K with $B\parallelsum c$, when the large MR background is removed. Fast Fourier transform (FFT) of the oscillatory resistivity between $(14~{\rm T})^{-1}$ and $(8~{\rm T})^{-1}$ reveals well-defined peaks, indicating the detection of quantum oscillations due to Landau quantizations (Fig.~\ref{fig3}(a)). We successfully identify all frequencies predicted by the DFT calculation, to be discussed below in relation to the angular dependence of the SdH frequencies. No splitting of the peaks was observed, in contrast to the case of TmSb and CeSb \cite{Wang2018, Ye2018}, consistent with the non-magnetic and the single-domain nature of our crystal. With $B\parallelsum c$, the SdH spectrum contains $\alpha_1$ and $\alpha_2$, coming from the electron ellipsoids located at the X point of the Brillouin zone; and $\beta$, $\gamma$, and $\zeta$, three hole pockets centred at $\Gamma$ point. Additionally, harmonics of these fundamental frequencies are detected. The amplitudes of these frequencies are sensitive to temperature, allowing the determination of the effective masses via the thermal damping factor $R_T$ of the Lifshitz-Kosevich (LK) theory:
\begin{equation}
R_T=\frac{14.69\ m^*T/B}{\sinh (14.69\ m^*T/B)}.
\label{LKeq}
\end{equation}
Figure~\ref{fig3}(b) displays the results of the fit to the temperature dependence of the amplitude. The effective masses obtained are tabulated in Table~1, along with the corresponding SdH frequencies when the field is along the c-axis. The effective masses range from 0.16$m_0$ to 0.61$m_0$, where $m_0$ is the rest mass of the electron. These effective masses are similar to those determined for YSb and LaSb \cite{Tafti2016A, Yu2017}.

\begin{figure}[!t]\centering
       \resizebox{8cm}{!}{
              \includegraphics{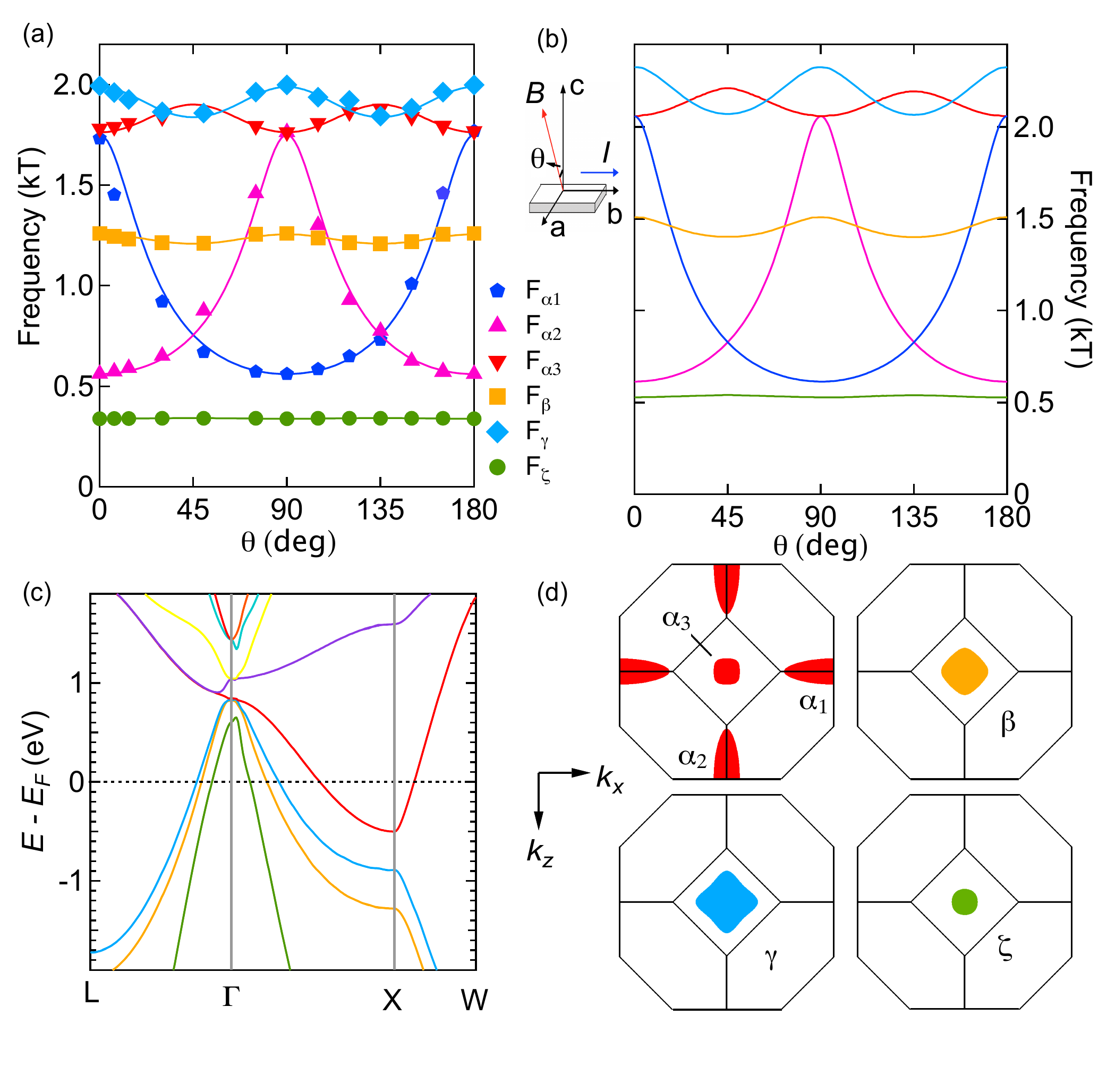}}
              \caption{\label{fig4} (a) Angular dependence of the measured SdH frequencies. For clarity, only the fundamental frequencies are displayed. (b) Calculated angular dependence of the quantum oscillation frequency via DFT calculations. The definition of the angle $\theta$ is shown. (c) Energy-momentum dispersion ($E({\bf k})$) along selected high symmetry directions. (d) The Fermi surfaces extracted from DFT calculations, showing three hole pockets ($\zeta$, $\beta$, and $\gamma$) and one set of symmetrically equivalent electron ellipsoids ($\alpha_1$, $\alpha_2$ and $\alpha_3$).}
\end{figure}

We now construct the full Fermi surface topology from the SdH data. Figure~\ref{fig4}(a) plots the angular dependence of SdH frequencies $F(\theta)$ (symbols). A slightly more elaborated version of Fig.~\ref{fig4}(a), which shows the SdH spectra at different $\theta$, can be found in the Supplemental Material \cite{SUPP}. We define $\theta$ as the angle between the $c$-axis and the field direction in the $ac$-plane (see Fig.~\ref{fig4}). For clarity, only the fundamental frequencies are shown in Fig.~\ref{fig4}(a).  
Figure~\ref{fig4}(b) shows the theoretical quantum oscillation frequencies extracted from the Fermi surfaces calculated by DFT as illustrated in Fig.~\ref{fig4}(d). Although the calculated frequencies are slightly different from the experimental values, the angular dependences are in excellent agreement. In addition to the branches with strong angular dependences, the remaining branches $\zeta$, $\beta$ and $\gamma$ in Fig.~\ref{fig4}(a) all have instantly recognizable correspondence with the calculated $F(\theta)$. Together with the energy-momentum dispersion ($E({\bf k})$) displayed in Fig.~\ref{fig4}(c), we identify $\zeta$, $\beta$ and $\gamma$ as the hole pockets centred at $\Gamma$, and $\alpha_1$, $\alpha_2$ and $\alpha_3$ as the electron ellipsoids centered at the X point of the Brillouin zone. Note that the existence of three $E({\bf k})$ branches in the vicinity of $\Gamma$, resulting in three hole pockets, is very clear in our calculations.

With an increasing $\theta$, $F_{\alpha_2}$ increases and reaches a maximum value at $\theta=90^\circ$ (Fig.~\ref{fig4}(a)). This is consistent with the fact that $\alpha_2$ is a closed, ellipsoidal pocket. $F_{\alpha_1}$ exhibits a similar angular dependence, except for a $90^\circ$ offset relative to $F_{\alpha_2}$. Thus, we assign them as electron pockets with the long-axis in the field rotation plane ($k_x$-$k_z$ plane), and they are orthogonal to each other. The angular dependence of SdH frequencies can be quantitatively described by:
\begin{equation}
F_{\alpha_1}(\theta)= \frac{F^{\rm min}_{\alpha_1} }{\sqrt{\cos^2(\theta-90^{\circ}) + \lambda_{\alpha_1}^{-2} \sin^2(\theta-90^{\circ})}}, 
\label{eq:alpha1}
\end{equation}
\begin{equation}
F_{\alpha_2}(\theta)= \frac{F^{\rm min}_{\alpha_2} }{\sqrt{\cos^2(\theta) + \lambda_{\alpha_2}^{-2} \sin^2(\theta)}}, 
\label{eq:alpha2}
\end{equation}
where $F_i^{\rm min}$ is the minimum frequency and $\lambda_i$ is the anisotropy factor for pocket $i$, 
For an ellipsoid,  $\lambda_i$ can be written as $k_F^L/k_F^S$, the ratio of the long semi-axis to the short semi-axis. From the fitting (solid lines in Fig.~\ref{fig4}(a)), $(F^{\rm min}_{\alpha_1}, \lambda_{\alpha_1})$=(561~T, 3.15) and ($F^{\rm min}_{\alpha_2}, \lambda_{\alpha_2})$=(561~T, 3.14). These results again show that the two ellipsoids are identical in size and shape, consistent with the calculated Fermi surfaces displayed in Fig.~\ref{fig4}(d).

The calculation shows that there is one more electron pocket $\alpha_3$, which is also identical to $\alpha_1$ and $\alpha_2$ but its long axis is perpendicular to the $k_x$-$k_z$ plane. In other related monopnictides, it has been challenging to detect this particular pocket \cite{Xu2017,Han2017,Sun2016,Kumar2016}. However, we do not face this difficulty here: the third electron pocket $\alpha_3$ is clear in our SdH data (down triangles in Fig.~\ref{fig4}(a)).   
Note that the projection of $\alpha_3$ on the $k_x$-$k_z$ plane is not circular but square-like, with two of its sides parallel to $\Gamma$--X. Therefore, the frequency coming from $\alpha_3$ should be slightly anisotropic: the minima of $F_{\alpha_3}$ must coincide with the maxima of $F_{\alpha_1}$, $F_{\alpha_2}$.  With these considerations, we describe $F_{\alpha_3}$ with the following equation which possesses a four-fold symmetry:
 \begin{equation}
F_{\alpha_3}(\theta)= \frac{F^{\rm min}_{\alpha_3} }{\sqrt{\cos^2(2\theta-180^{\circ}) + \lambda_{\alpha_3}^{-2} \sin^2(2\theta-180^{\circ})}}. 
\label{eq:alpha3}
\end{equation}
Here, the fitting of the data to Equation~\ref{eq:alpha3} gives $F^{\rm min}_{\alpha_3}= 1763~$T and $\lambda=1.07$. Note that the value of $F^{\rm min}_{\alpha_3}$ is in excellent agreement with $F_{\alpha_1}(0^{\circ})=1767$~T and $F_{\alpha_2}(90^{\circ})=1762$~T, as expected.

The projections of $\beta$ and $\gamma$ pockets on the field rotation plane are anisotropic with a four-fold symmetry. Their diagonals are along $\Gamma$--X, so the maximal frequencies are located at $90^\circ$ and $180^\circ$. The angle dependence of $F_\beta$, $F_\gamma$ can be fitted with:
\begin{equation}
F_{\beta/\gamma}(\theta)= \frac{F^{\rm min}_{\beta/\gamma} }{\sqrt{\cos^2(2\theta-90^{\circ}) + \lambda_{\beta/\gamma}^{-2} \sin^2(2\theta-90^{\circ})} }
\label{eq:beta}
\end{equation}
The fitting of the SdH data gives $F^{\rm min}_{\beta}=1209$~T, $\lambda_{\beta}=1.04$, and $F^{\rm min}_{\gamma} =1844$~T, $\lambda_{\gamma}=1.08$. Finally, the frequency for the smallest hole pocket $\zeta$ is $\sim$~340~T with a negligible anisotropy. Therefore, we successfully identify all Fermi pockets predicted in DFT calculations, and the SdH data further enable us to describe their shape and size accurately.

Using the Onsager relation, $F=(\Phi_0/2\pi^2)A$, we can estimate the Fermi wavevectors and the volume of all Fermi pockets, where $A$ is the area enclosed by the extremal orbit perpendicular to the magnetic field and $\Phi_0$ is the flux quantum. For ellipsoidal electron pockets, the short semi-axis $k_F^S$ can be estimated by $A_S=\pi {k_F^S}^2$ while the long semi-axis can be calculated by $A_L= \pi k_F^S k_F^L$, where $A_S$ and $A_L$ are the area associated with the minimum and maximum in $F_{\alpha_1}$, respectively. Note that we have approximated $A_S$ as a circle.
The hole pockets are approximated as spheres. These approximations are justifiable because $\lambda$ is close to unity for these cases. From these volumes, we can calculate the carrier densities: the total electron density $n$, coming from three ellipsoids, is $7.099\times10^{20}~{\rm cm}^{-3}$, and the hole density of $\beta$, $\gamma$ and $\zeta$ is $2.458\times10^{20}~{\rm cm}^{-3}$, $4.800\times10^{20}~{\rm cm}^{-3}$ and $3.556\times10^{19}~{\rm cm}^{-3}$, respectively. Therefore, the total hole density $p$ is $7.613\times10^{20}~{\rm cm}^{-3}$, giving the ratio $n/p\approx0.93$.

Arming with the full knowledge of the Fermi surface, we now revisit the topic of the large MR. Our calculated $E({\bf k})$ in Fig.~\ref{fig4}(c) shows that the Sc $d$ states and Sb $p$ states are separated by 0.39~eV at X. The existence of such a large gap is robust, even when different functionals are used \cite{SUPP}. In the absence of a band crossing, ScSb is topologically trivial. Therefore, mechanisms such as the field-induced breakdown of topological protection cannot explain the large MR here \cite{Liang2015}. A two-band model was employed to describe the magnetoresistance ${\rm MR}=n p \mu_e\mu_h(\mu_e+\mu_h)^2B^2 / [(n\mu_e+p\mu_h)^2+(n-p)^2(\mu_e\mu_h)^2B^2]$, where $\mu_e$ and $\mu_h$ are mobilities for electrons and holes, respectively. We use $n$ and $p$ determined from SdH oscillations. Although $\mu_e$ and $\mu_h$ cannot be determined individually due to the nearly perfect compensation of electron and hole ($n \simeq p$) and the absence of Hall data, we can extract an effective mobility $\mu_{\rm eff} = \sqrt{\mu_e\mu_h}$ at 2~K, which is $1.25(\pm0.04)\times10^4~{\rm cm}^2{\rm (Vs)}^{-1}$. Moreover, we found that a significant difference between electron and hole mobilities is essential to fit the experimental MR \cite{SUPP}. Previously, mobilities have also been extracted from two-band model for YSb ($\mu_e=0.935\times10^4~{\rm cm}^2{\rm (Vs)}^{-1}$, $\mu_h=1.056\times10^4~{\rm cm}^2{\rm (Vs)}^{-1}$)\cite{Xu2017} and LaSb ($\mu_e=1.118\times10^4~{\rm cm}^2{\rm (Vs)}^{-1}$, $\mu_h=0.964\times10^4~{\rm cm}^2{\rm (Vs)}^{-1}$)\cite{Han2017}. From these mobilities, $\mu_{\rm eff}$ can be calculated to be $1.021\times10^4~{\rm cm}^2{\rm (Vs)}^{-1}$ and $1.019\times10^4~{\rm cm}^2{\rm (Vs)}^{-1}$, for YSb and LaSb, respectively. These values are close to $\mu_{\rm eff}$ extracted for ScSb. These observations indicate that a nearly perfect electron-hole compensation together with mobility mismatch can explain the non-saturating XMR behaviour in ScSb, similar to the cases of YSb \cite{Xu2017} and LaSb \cite{Han2017}, despite the fact that ScSb has an additional hole pocket. 

\section{IV. Conclusions}

In summary, we have measured the magnetoresistance of ScSb. The magnetoresistance is nearly quadratic in field and non-saturating at 14~T. Kohler scaling is obeyed over a large temperature range. Clear Shubnikov-de Haas oscillations have been detected at various angles, allowing the full construction of the  Fermi surface. All bulk Fermi pockets predicted by bandstructure calculations are identified experimentally. Calculations show the absence of band inversion at X, and hence ScSb is topologically trivial. The large magnetoresistance in ScSb is attributed to a nearly perfect compensation of electron and hole, possibly accompanied by mobility mismatch, indicating its semiclassical origin.

\section{Acknowledgments}
\begin{acknowledgments}
We acknowledge technical support from Yuet Ching Chan and King Yau Yip, and financial support from
Research Grants Council of Hong Kong (ECS/24300214, ECS/24300814, GRF/14301316, GRF/14300117),
CUHK Direct Grant (4053223, 4053299), National Natural Science
Foundation of China (11504310) and 
Ministry of Science and Technology of Taiwan (MOST-106-2112-M-006-013-MY3)

$^\ddagger$Y.J.H. and E.I.P.A. contributed equally to this work. 
\end{acknowledgments}



\end{document}